\documentclass[Physsubmission, Phys]{SciPost}

\hypersetup{
    colorlinks,
    linkcolor={red!50!black},
    citecolor={blue!50!black},
    urlcolor={blue!80!black}
}

\usepackage{subfig}

\usepackage[bitstream-charter]{mathdesign}
\DeclareSymbolFont{usualmathcal}{OMS}{cmsy}{m}{n}
\DeclareSymbolFontAlphabet{\mathcal}{usualmathcal}

\begin{document}


\begin{center}{\Large \textbf{
Helix string fragmentation and charged particle correlations  \\ with
ATLAS \\
}}\end{center}

\begin{center}
\v{S}\'{a}rka Todorova-Nov\'{a} \textsuperscript{$\star$} 
on behalf of the ATLAS Collaboration \footnote{Copyright 2022 CERN for the benefit of the ATLAS Collaboration. Reproduction of this article or parts of it is allowed as specified in the CC-BY-4.0 license.}
\end{center}

\begin{center}
{\bf } IPNP, Charles University, Prague
\\
* sarka.todorova@cern.ch
\end{center}

\begin{center}
\today
\end{center}


\definecolor{palegray}{gray}{0.95}
\begin{center}
\colorbox{palegray}{
  \begin{tabular}{rr}
  \begin{minipage}{0.1\textwidth}
    \includegraphics[width=23mm]{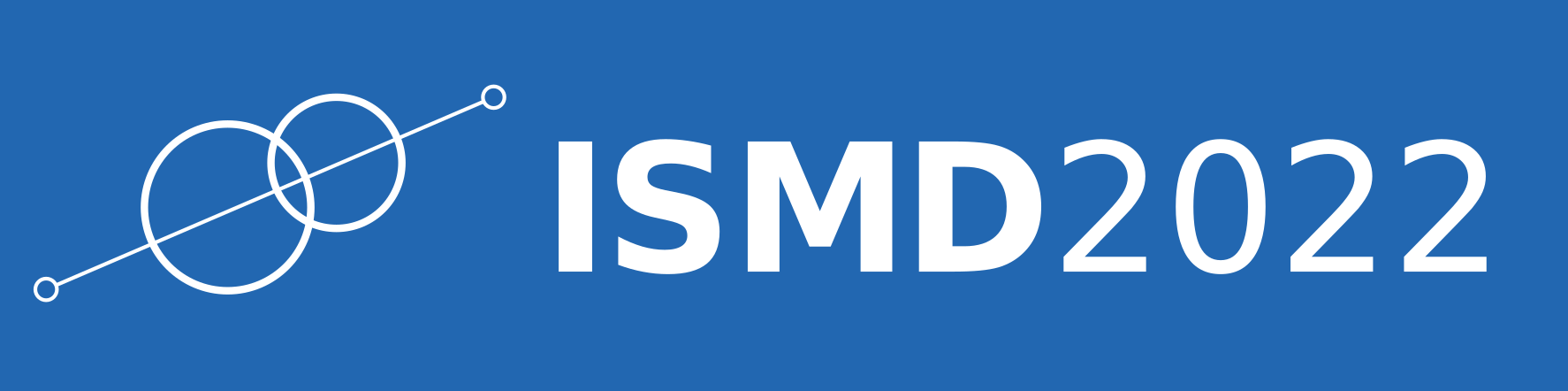}
  \end{minipage}
  &
  \begin{minipage}{0.8\textwidth}
    \begin{center}
    {\it 51st International Symposium on Multiparticle Dynamics (ISMD2022)}\\ 
    {\it Pitlochry, Scottish Highlands, 1-5 August 2022} \\
    \doi{10.21468/SciPostPhysProc.?}\\
    \end{center}
  \end{minipage}
\end{tabular}
}
\end{center}

\section*{Abstract}
{\bf
Correlations between charged particles provide important insight into the hadronization process. The
analysis of the momentum difference between charged hadrons in $pp$, $p$-lead, and lead-lead collisions
at LHC is performed by the ATLAS Collaboration in order to study the dynamics of hadron formation. The spectra
of correlated hadron chains are explored and compared to the predictions based on the quantized
fragmentation of a three dimensional QCD helix string. This provides an alternative view of the
two-particle correlation phenomenon typically attributed to the Bose-Einstein interference.
}

\vspace{10pt}

\section{Introduction}

\begin{figure}[t]
\centering
 \includegraphics[width=0.42\textwidth]{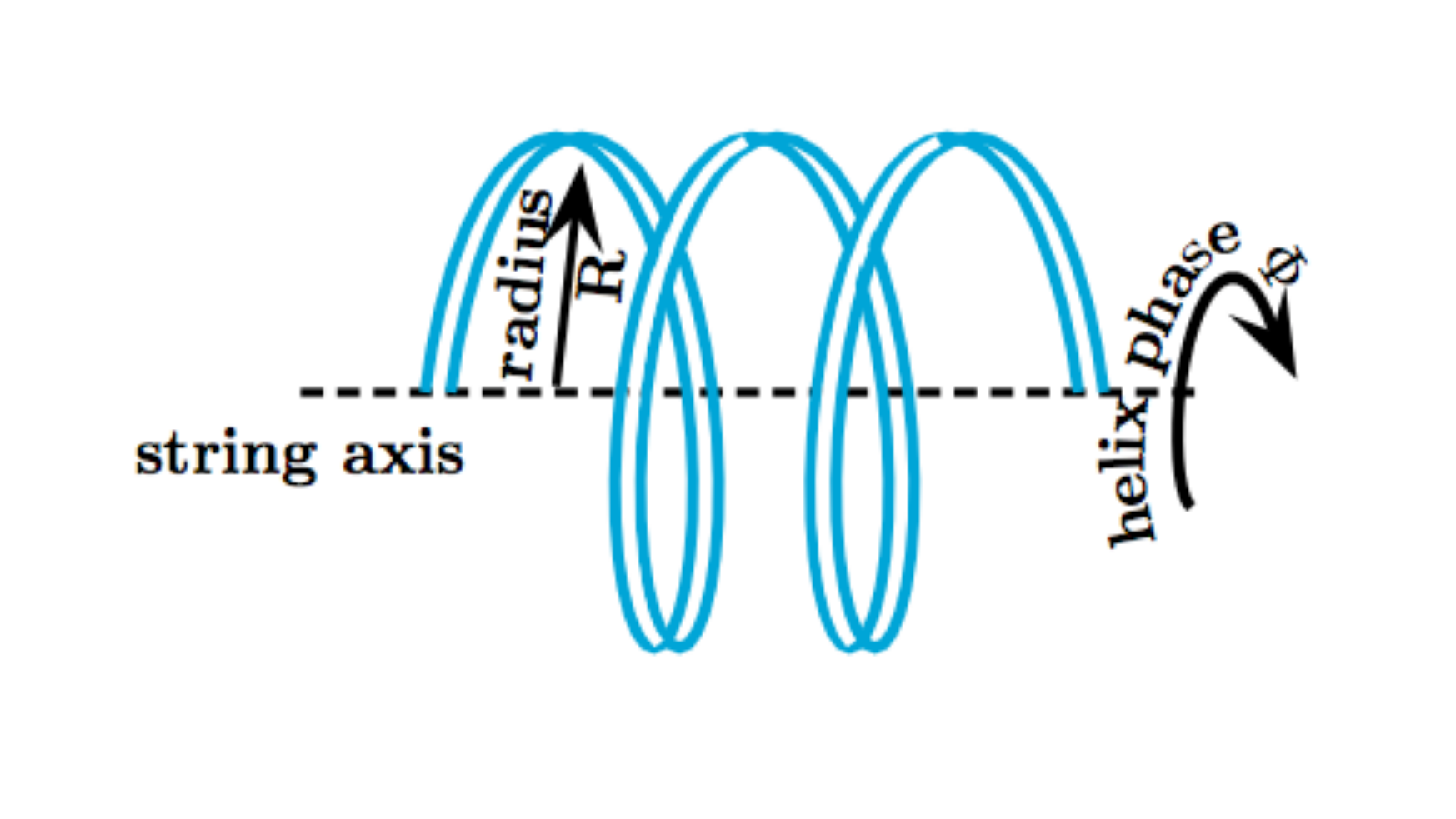}
 \includegraphics[width=0.57\textwidth]{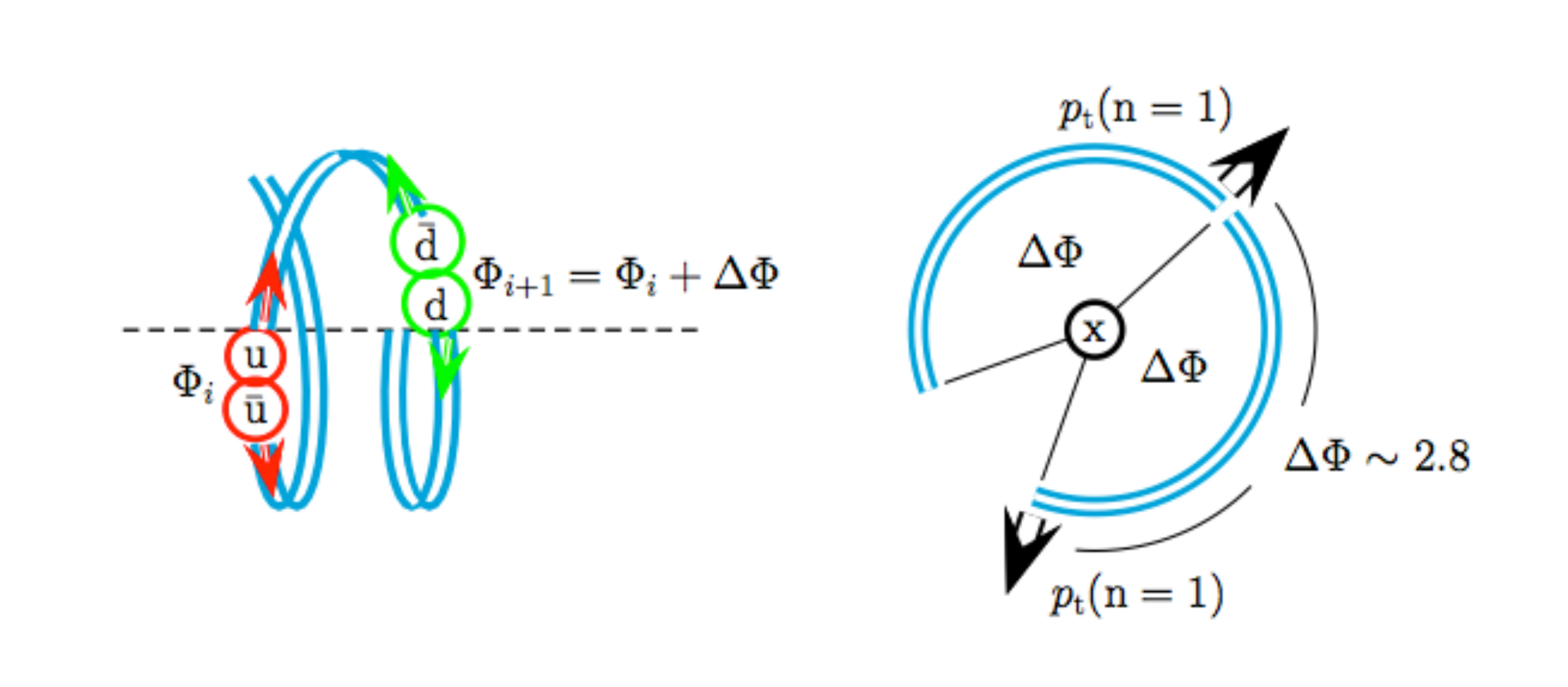}
\caption{ Parametrization of the helical QCD string (left) and of the quantized string breakup (middle),  which creates correlation between colour-adjacent direct pions (right). 
 \label{fig:helix}}
\end{figure}

    The notion that correlations between hadrons are an important source of information about hadron formation is widely accepted but not sufficiently explored.  Recently,  an alternative approach to the modeling of the confinement, replacing the 1-dimensional string of the Lund fragmentation model by a 3-dimensional helical string, brought some insight into the rules governing the hadronization. There are several essential properties the 3-dimensional string brings into consideration: the intrinsic transverse momentum of hadron is defined by the transverse shape of the string as well as the correlation between intrinsic transverse and longitudinal momentum components.  Of special interest is the possibility of studying, for the first time, the causal relations between string breakups which define direct hadrons. The requirement of a cross-talk between endpoint vertices reveals a quantization scheme in which different hadron species correspond to a helical string breaking in regular intervals of helix phase $\Delta\Phi\sim$2.8~rad \cite{helix2,baryons}. The mass spectrum of light hadrons is described, with a precision of 1-3\%, with help of only two parameters: $\Delta\Phi$ and the energy scale $\kappa R$, where  $\kappa$ stands for the string tension and $R$ for the radius of the helix (see Figure~\ref{fig:helix}).  Since the quantization proceeds in the transverse mass $m_t = \sqrt{m^2 + p_t^2}$ rather than mass $m$ alone, the model predicts non-trivial correlations between colour-adjacent hadrons. Two studies of these correlations were performed by the ATLAS Collaboration\cite{ATLAS} using Run 1 LHC data (\cite{ATLAS-AO},\cite{ATLAS-chains}).

\section{Anomalous production of like-sign pions}
  For the specific case of a helical string with a constant string pitch, fragmenting into a chain of direct pions, the momentum difference $Q=\sqrt{-(p_i-p_j)^2}$ between pions can be calculated as a function of their rank difference, where the rank describes the hadron ordering along the string according to colour flow. An anomaly appears for colour adjacent pairs (of rank difference 1) which have the largest momentum difference ($\sim$ 270 MeV), while pairs with rank difference 2 are separated by $\sim$ 90 MeV only. This is of course a consequence of $\Delta\Phi$ being $\sim$ 2.8 rad, and colour adjacent pion pairs nearly back-to-back in the transverse plane. This anomaly is coupled with a charge-combination asymmetry since local charge conservation in string breaking forbids creation of colour-adjacent pairs with identical sign of charge.  Another prediction of the model suggests the mass of a chain of $n$ direct pions has the smallest mass possible, among any combination of $n$ hadrons produced from a given piece of string.

\begin{figure}[h]
\centering
\subfloat[]{  \includegraphics[width=0.49\textwidth]{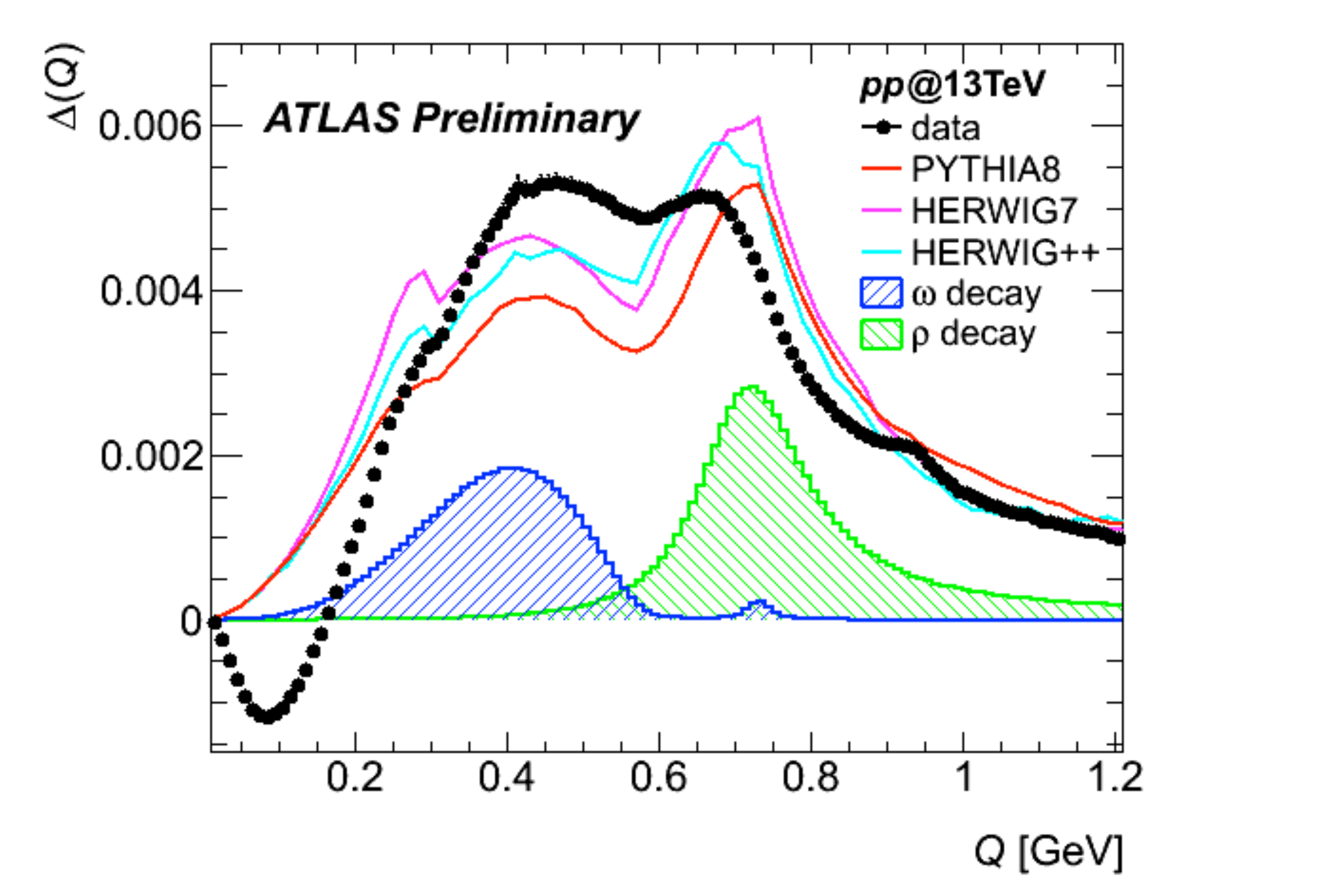} \label{fig:chains_a} }
\subfloat[]{ \includegraphics[width=0.49\textwidth]{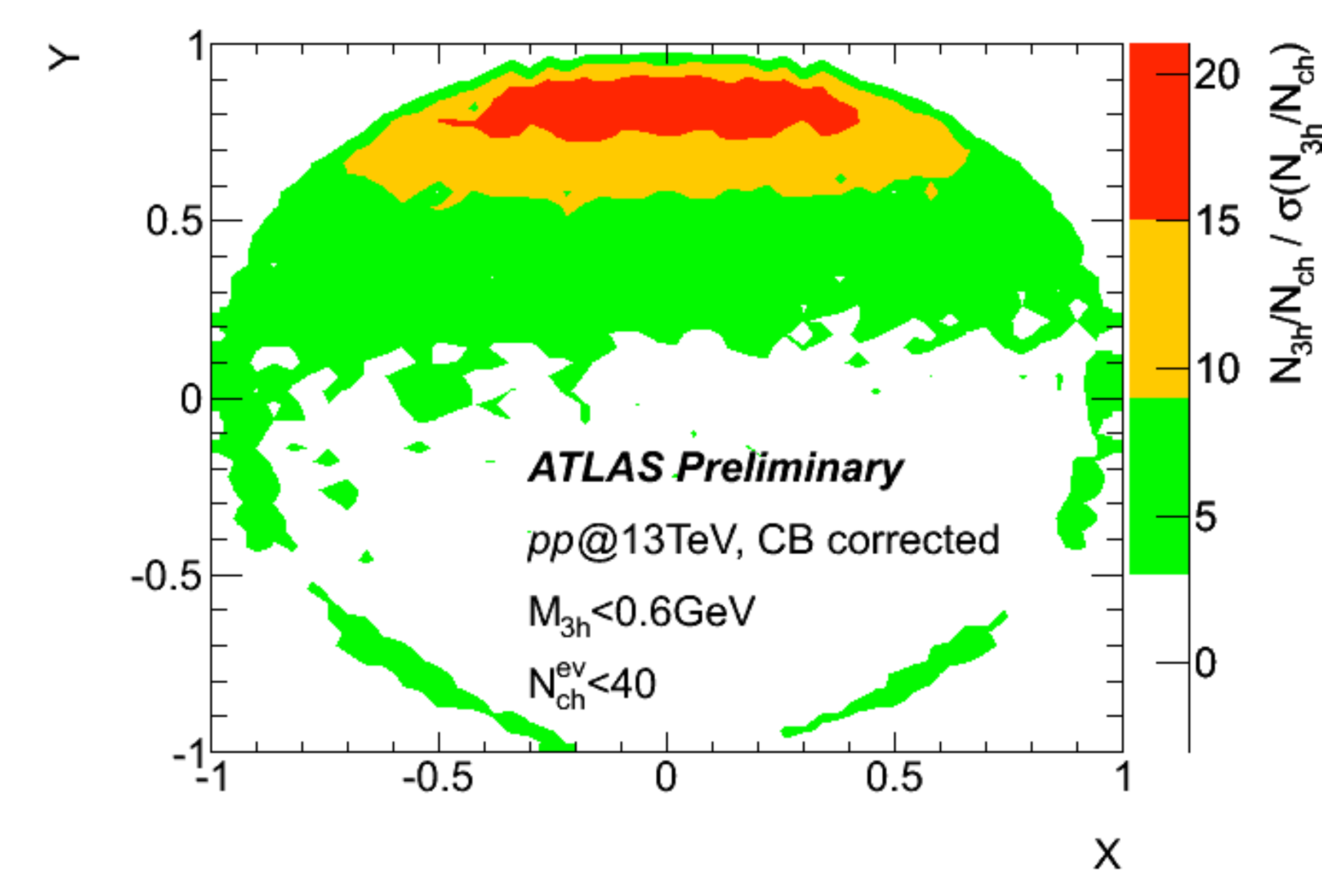}  \label{fig:chains_b} } \\
\subfloat[]{ \includegraphics[width=0.49\textwidth]{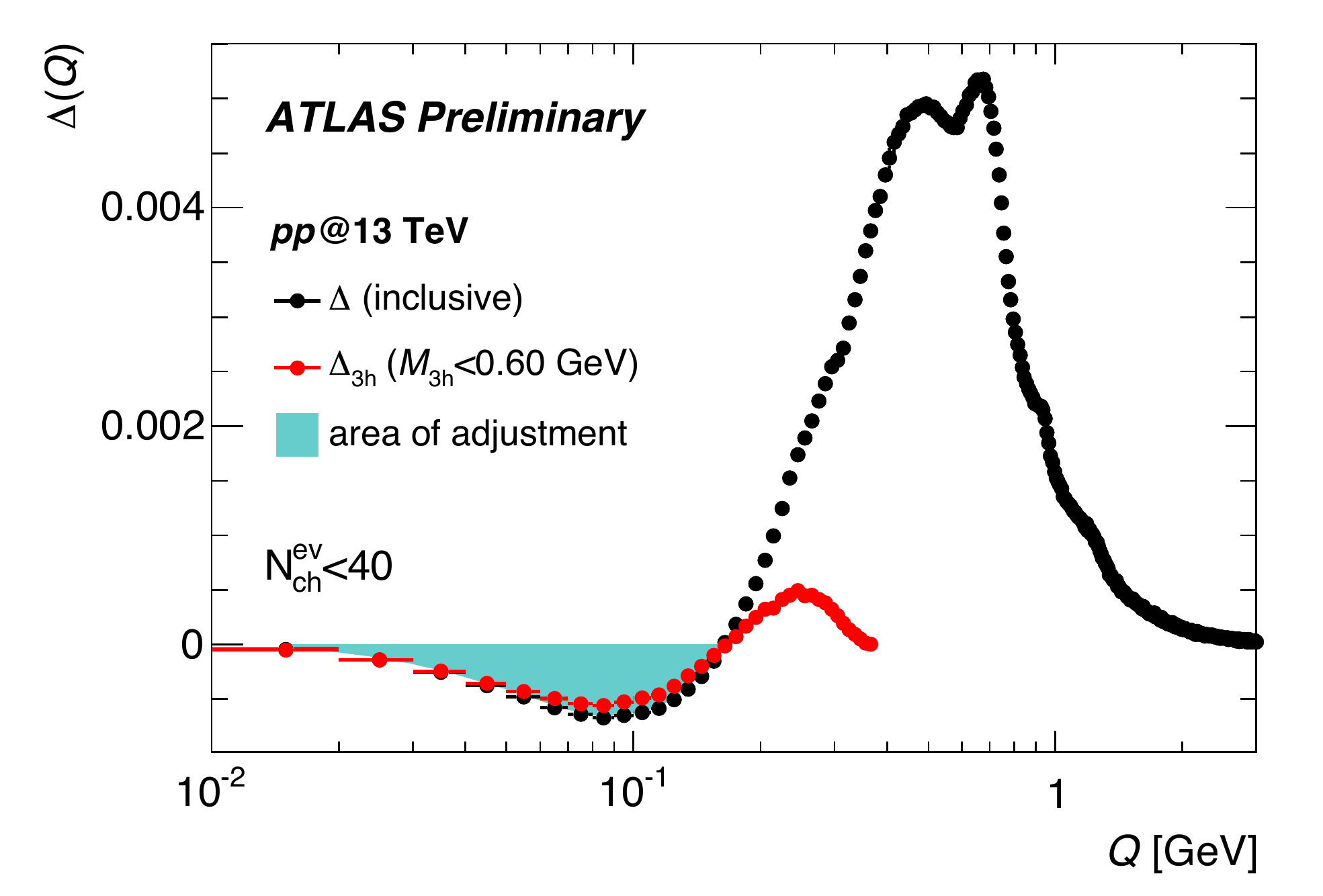} \label{fig:chains_c} }
\subfloat[]{ \includegraphics[width=0.49\textwidth]{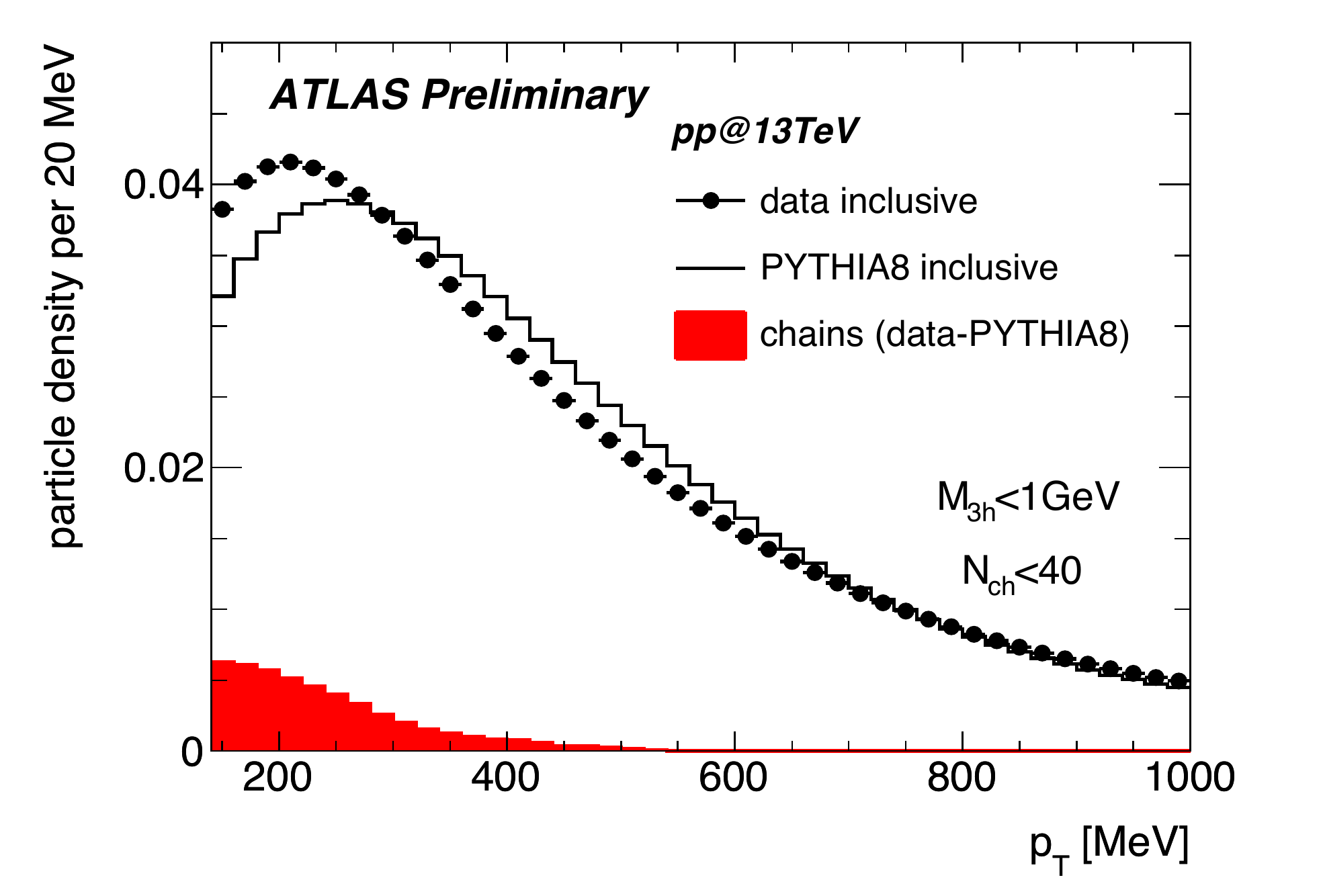} \label{fig:chains_d} }
\caption{Anomalous production of like-sign hadrons (pions) seen as a negative part of measured $\Delta(Q)$ (a)  is explained by the presence of mass-minimized, charge ordered triplets ("chains") in the configuration predicted by quantized fragmentation  as shown in the Dalitz plot (b). The bottom plots
indicate the contribution ($\Delta_{3h}$) of the chain selection to $\Delta(Q)$(c) and to the inclusive $p_\mathrm{T}$ spectrum (d). Source: Ref~\cite{confnote}.}
\label{fig:chains}
\end{figure}

  Model predictions are tested using Run 2 ATLAS data, Ref.\cite{confnote}, with the help of the observable \\
 $\Delta(Q)= \frac{1}{N_\mathrm{ch}}  [ N^{OS}(Q) - N^{LS}(Q) ]$ based on the difference between inclusive spectra of like-sign (LS) and opposite-sign (OS) hadron pairs. $\Delta(Q)$ is uniquely sensitive to the momentum difference between colour-adjacent hadrons\cite{ismd22_cr}. $N_\mathrm{ch}$ indicates the number of particles in the sample. Anomalous production of LS pairs appears as a negative part of the $\Delta(Q)$ distribution, Figure \ref{fig:chains}\subref{fig:chains_a}. The discrepancies observed between data and conventional hadronization models are substantial which means the modeling does not guarantee a reliable estimation of hadronization systematics. 
In a model independent way, the excess of LS pairs is associated with the presence of charge-ordered triplets (chains) with properties predicted by the quantized fragmentation (Figure \ref{fig:chains}\subref{fig:chains_b},\subref{fig:chains_c}), and with the modification of the inclusive $p_\mathrm{T}$ spectra, Figure \ref{fig:chains}\subref{fig:chains_d}. The latter does not appear by surprise either, since the intrinsic $p_t$ of the direct pions is $\sim$130 MeV in the model and the strings tend to be aligned with the beam direction in minimum bias samples.

\begin{figure}[h]
\centering
\subfloat[]{  \includegraphics[width=0.49\textwidth]{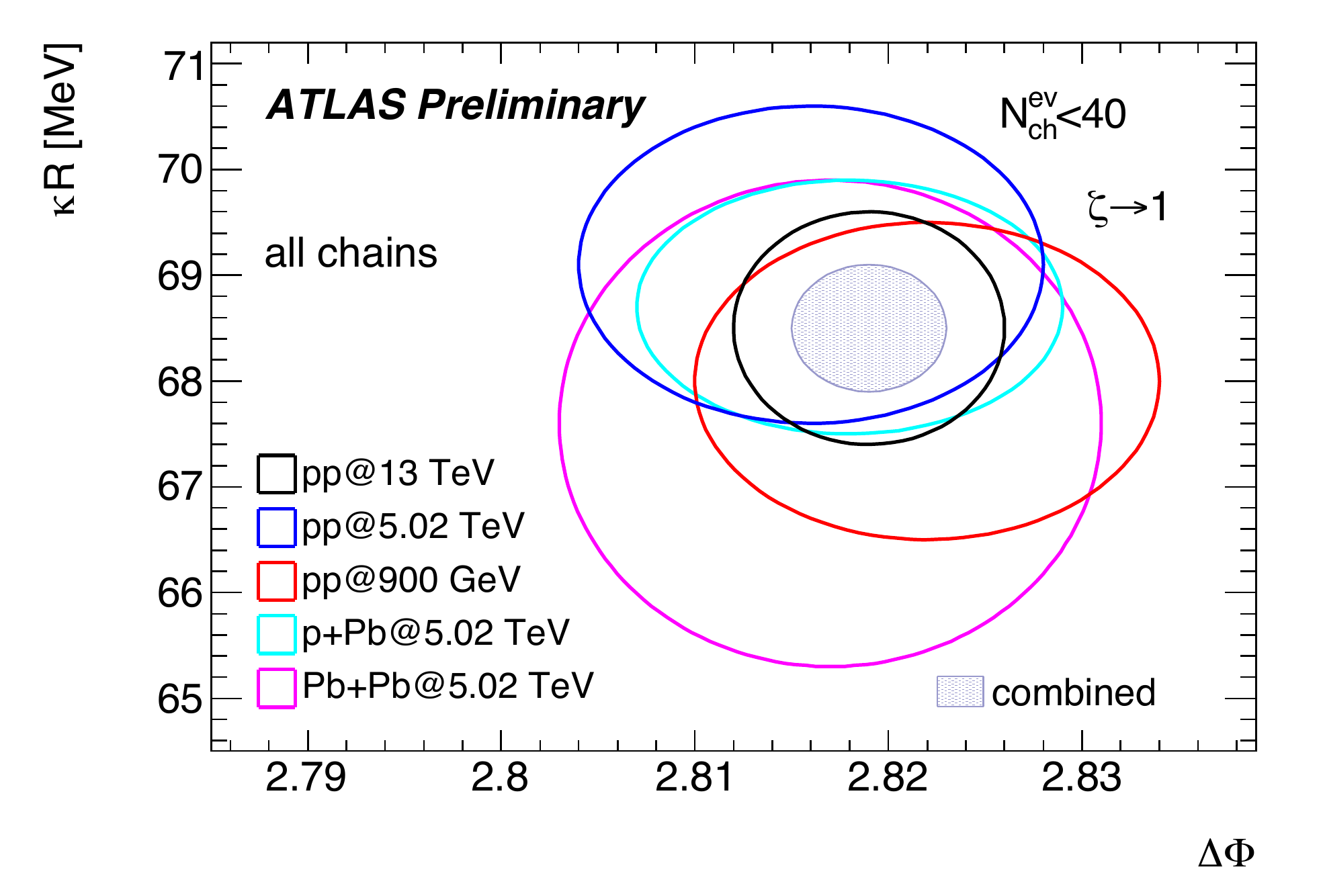} \label{fig:params_a} }
\subfloat[]{  \includegraphics[width=0.49\textwidth]{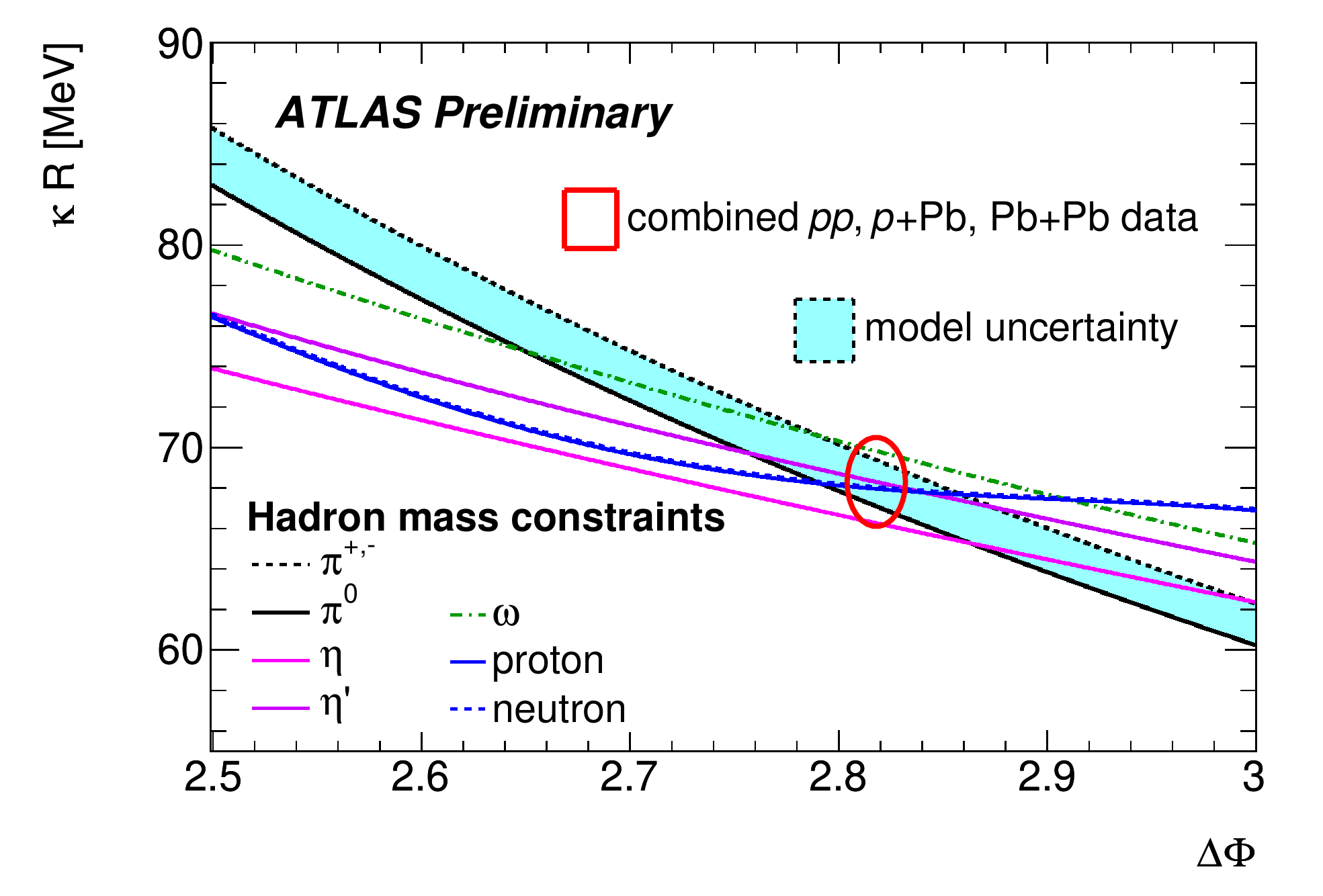} \label{fig:params_b} } 
\caption{Measurement of string parameters using particle correlations. (a) Comparison of intermediate results obtained in various Run 2 samples. (b)  Combined final result compared with constraints derived from hadron masses. The blue area indicates the intrinsic uncertainty of the model of quantized fragmentation. Source: Ref~\cite{confnote}.}
\label{fig:params}
\end{figure}

 The shape of $\Delta_{3h}$ (Figure \ref{fig:chains}\subref{fig:chains_c}) is used to perform an independent measurement of string parameters, in $pp$ data collected at various collision energies (0.9, 5 and 13 TeV) as well as in data from heavy-ion collisions ($p$+Pb and Pb+Pb). The measurement is done in a way which eliminates the contribution from (the unknown) variation of longitudinal shape of the string, with the help of an additional observable sensitive to the local variation of the fragmentation function, $\zeta=min(\frac{|p_i|}{|p_j|},\frac{|p_j|}{|p_i|})$. Very good agreement is obtained between all samples, Figure \ref{fig:params}\subref{fig:params_a}. The combined result is shown in Figure \ref{fig:params}\subref{fig:params_b}  and compared with constraints imposed on the model by hadron masses.

\clearpage

\section{Signature of long pion chains}
\begin{figure}[h]
\centering
\subfloat[]{  \includegraphics[width=0.49\textwidth]{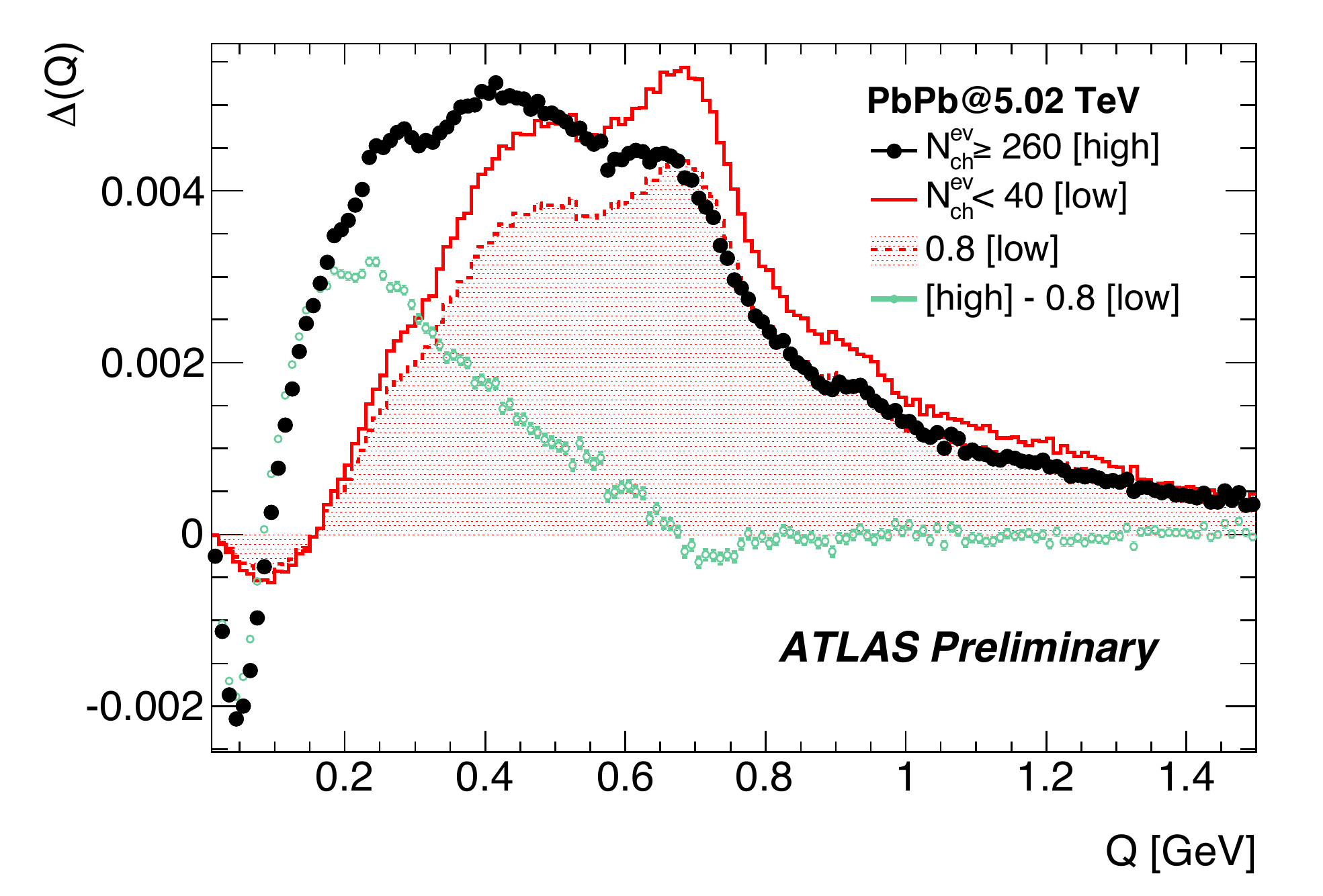} \label{fig:stripes_a} }
\subfloat[]{ \includegraphics[width=0.49\textwidth]{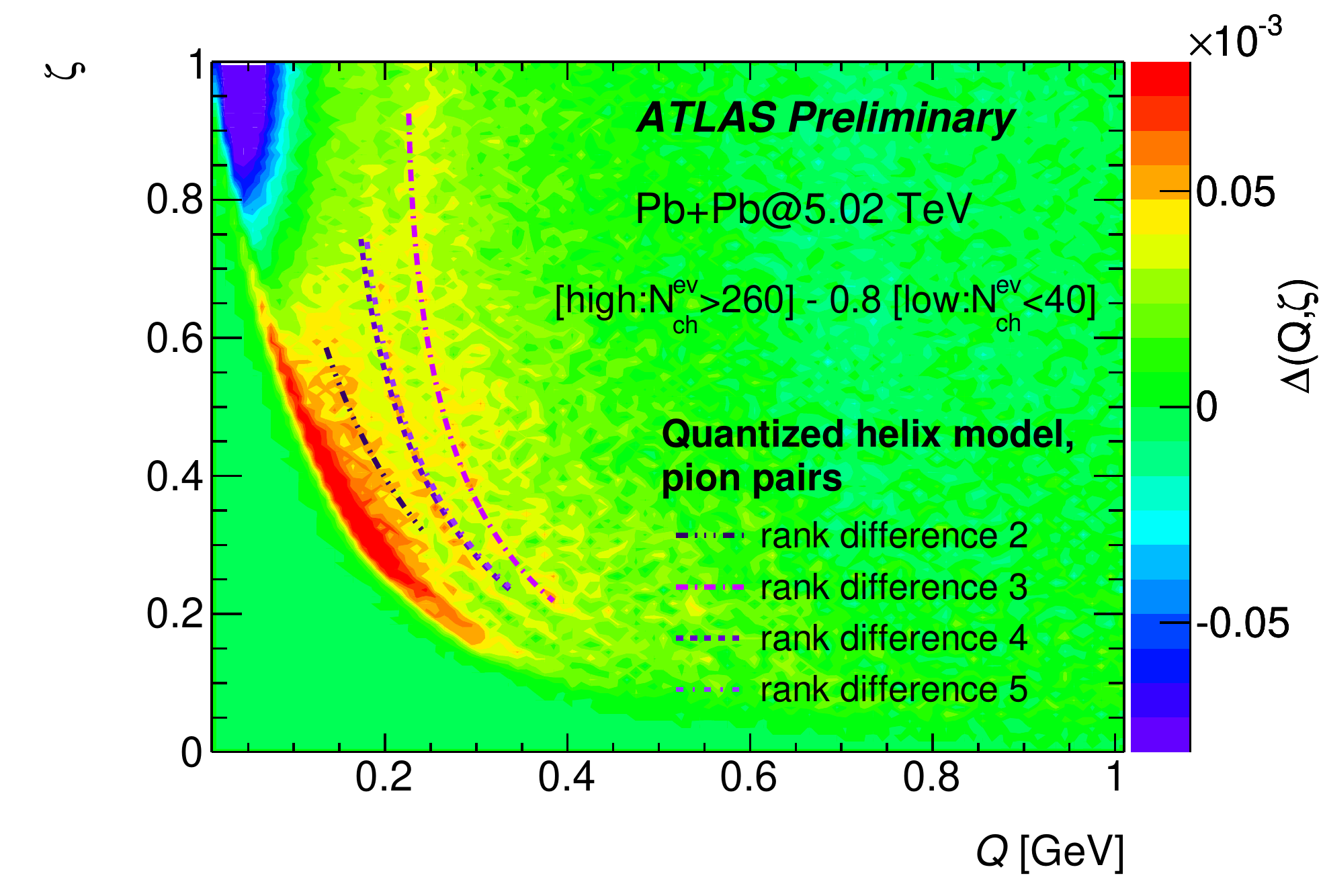} \label{fig:stripes_b} } 
\caption{ (a) The comparison of the shape of $\Delta(Q)$ in events with low and high charged particle multiplicity reveals an enhanced activity in the low-$Q$ region (green points) in peripheral high-multiplicity Pb+Pb collisions. (b) The enhanced soft signal in high multiplicity peripheral Pb+Pb events exhibits the signature of quantized long pion chains. Source: Ref~\cite{confnote}. }
\label{fig:stripes}
\end{figure}

 The shape of $\Delta(Q)$ evolves with particle multiplicity and this evolution indicates a growing contribution from hadronic sources with mass below 0.7 GeV, kinematically compatible with a signature of wounded nucleons, Figure \ref{fig:stripes}\subref{fig:stripes_a}. A two-dimensional study ($Q,\zeta$) of this low-mass contribution in peripheral Pb+Pb collisions reveals a distinct stripe structure which fits the expected quantized signature of direct pion pairs from long pions chains, Figure \ref{fig:stripes}\subref{fig:stripes_b}. The observation of long pions chains may be instrumental in understanding of long-range correlations.

\section{Conclusion}
  The anomalous production of like-sign charged pions is studied using LHC Run~2 data and found to conform to predictions of the model of quantized fragmentation which is - at the time of writing - the only model capable of calculating the mass spectrum of light hadrons from properties of the parametrized strong field. 
Signature of quantized fragmentation is found by the ATLAS Collaboration in 1-, 2- and 3-particle spectra and the link between these phenomena is experimentally established in a model-independent way. Quantitave estimates suggest the quantized fragmentation
accounts for the entire anomalous production of close like-sign pairs traditionally attributed to Bose-Einstein interference.

\paragraph{Funding information}
This work was partially supported by the Inter-Excellence/Inter- Transfer grant LTT17018 and the Research Infrastructure project LM2018104 funded by Ministry of Education, Youth and Sports of the Czech Republic, and the Charles University project UNCE/SCI/013.





\end{document}